\RequirePackage[log]{snapshot}

\documentclass[%
 reprint,
 superscriptaddress,
 amsmath,amssymb,
 prb,
longbibliography
]{revtex4-1}

\usepackage{graphicx}
\usepackage{epstopdf}
\usepackage{dcolumn}
\usepackage{bm}
\usepackage{hyperref}
\hypersetup{
	colorlinks   = true, 
	urlcolor     = blue, 
	linkcolor    = blue, 
	citecolor    = blue 
}

\usepackage{color}

\usepackage{amsmath}
\usepackage[english]{babel}
\usepackage{graphicx}
\usepackage{tikz}
\usepackage{float}
\usepackage{subcaption}
\usepackage{ragged2e}
\usepackage{geometry}
\usepackage{wrapfig}
\usepackage{pdfpages}
\usepackage{xcolor}
\usepackage{natbib}

\makeatletter
\newcommand*{\defeq}{\mathrel{\rlap{%
                     \raisebox{0.3ex}{$\m@th\cdot$}}%
                     \raisebox{-0.3ex}{$\m@th\cdot$}}%
                     =}
\makeatother

\makeatletter
\AtBeginDocument{\let\LS@rot\@undefined}
\makeatother

\newcommand{\MgGd}{$\text{Mg}_{2}\text{Gd}_{3}\text{Sb}_{3}\text{O}_{14}$}

\setcitestyle{square,numbers}

\usepackage[normalem]{ulem}

\usepackage{color,soul}
\setul{0.5ex}{0.5ex}
\setulcolor{red}

\makeatletter
\newcommand\cwave[1][red]{\bgroup \markoverwith{\lower5\p@\hbox{\sixly \textcolor{#1}{\char58}}}\ULon}
\font\sixly=lasyb10 
\makeatother

\definecolor{mygray}{cmyk}{0, 0, 0, 0.3}
\definecolor{Ora}{cmyk}{0, 0.6, 0.8, 0.4}
\definecolor{cw_green}{cmyk}{0.14, 0, 0.90, 0.5}

\begin{document}

\title{Magnetic interactions in the tripod-kagome antiferromagnet $\text{Mg}_{2}\text{Gd}_{3}\text{Sb}_{3}\text{O}_{14}$ probed by static magnetometry and high-field ESR spectroscopy} 


\author{C. Wellm}
\affiliation{Leibniz IFW Dresden, 01069 Dresden, Germany}
\affiliation{Institute for Solid State and Materials Physics, TU Dresden, 01069 Dresden, Germany}

\author{J. Zeisner}
\affiliation{Leibniz IFW Dresden, 01069 Dresden, Germany}
\affiliation{Institute for Solid State and Materials Physics, TU Dresden, 01069 Dresden, Germany}

\author{A. Alfonsov}
\affiliation{Leibniz IFW Dresden, 01069 Dresden, Germany}

\author{M.-I. Sturza}
\affiliation{Leibniz IFW Dresden, 01069 Dresden, Germany}

\author{G. Bastien}
\affiliation{Leibniz IFW Dresden, 01069 Dresden, Germany}

\author{S. Ga\ss}
\affiliation{Leibniz IFW Dresden, 01069 Dresden, Germany}

\author{S. Wurmehl}
\affiliation{Leibniz IFW Dresden, 01069 Dresden, Germany}

\author{A. U. B. Wolter}
\affiliation{Leibniz IFW Dresden, 01069 Dresden, Germany}

\author{B. B\"uchner}
\affiliation{Leibniz IFW Dresden, 01069 Dresden, Germany}
\affiliation{Institute for Solid State and Materials Physics and W{\"u}rzburg-Dresden Cluster of Excellence ct.qmat, TU Dresden, D-01062 Dresden, Germany}

\author{V. Kataev}
\affiliation{Leibniz IFW Dresden, 01069 Dresden, Germany}

\begin{abstract}
We report an experimental study of the static magnetization $M(H,T)$ and high-field electron spin resonance (ESR) of polycrystalline \MgGd, a representative member of the newly discovered class of the so-called tripod-kagome antiferromagnets where the isotropic Gd$^{3+}$ spins ($S = 7/2$) form a two-dimensional kagome spin-frustrated lattice. It follows from the analysis of the low-$T$ $M(H)$-curves that the Gd$^{3+}$ spins are coupled by a small isotropic antiferromagnetic (AFM) exchange interaction $|J| \approx$ 0.3\,K.
The $M(H,T)$-dependences measured down to 0.5\,K evidence a long-range AFM order at $T_{\text{N}} = 1.7$\,K and its rapid suppression at higher fields $\geq 4$\,T.
ESR spectra measured in fields up to 15\,T are analyzed considering possible effects of demagnetizing fields, single-ion anisotropy and spin-spin correlations. While the demagnetization effects due to a large sample magnetization in high fields and its shape anisotropy become relevant at low temperatures, the broadening of the ESR line commencing already at $T\lesssim 30$\,K may indicate the onset of the spin-spin correlations far above the ordering temperature due to the geometrical spin frustration in this compound.
\end{abstract}

\date{\today}

\maketitle

\section{Introduction}

The archetypical two-dimensional (2D) kagome Heisenberg antiferromagnet is known to be significantly frustrated \cite{Yan2011}, the effects of which include suppression of magnetic order (with a spin-liquid ground state for $S$\,=\,1/2 \cite{Yan2011}) and a continuous evolution of physical parameters concomitant with the growth of short-range correlations rather than their sudden change due to phase transitions. In real materials, however, a 2D Heisenberg model is never ideally realized due to residual interlayer coupling as well as possible anisotropies of the spin-spin interactions. Those materials are therefore promising candidates for exciting and complex spin-related physics, where the competition of the different contributions may lift the ground state degeneracy, influence the frustration and the effective dimensionality of the system.

\begin{figure}[H]
	\begin{subfigure}[t]{0.22\textwidth}
		\begin{centering}
			\includegraphics[scale=0.2]{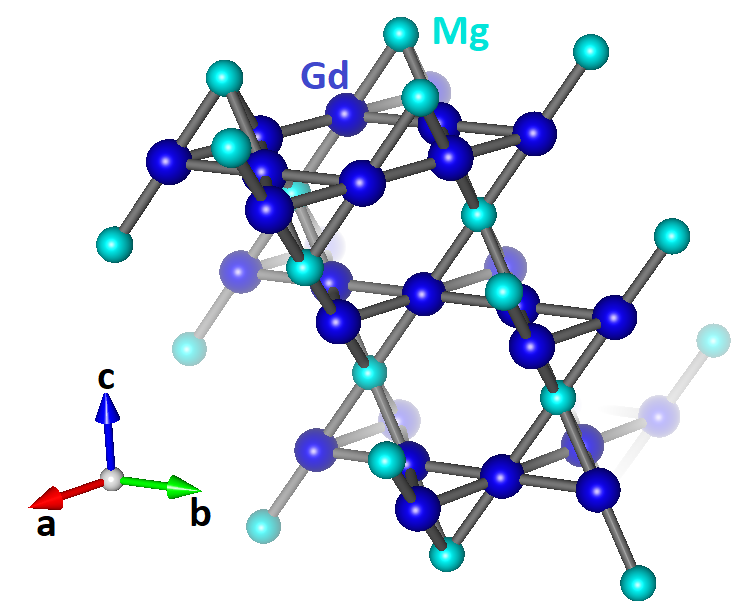}
			\caption{}
			\label{fig:pyrochlore}
		\end{centering}
	\end{subfigure}
	\begin{subfigure}[t]{0.22\textwidth}
		\begin{centering}
			\includegraphics[scale=0.15]{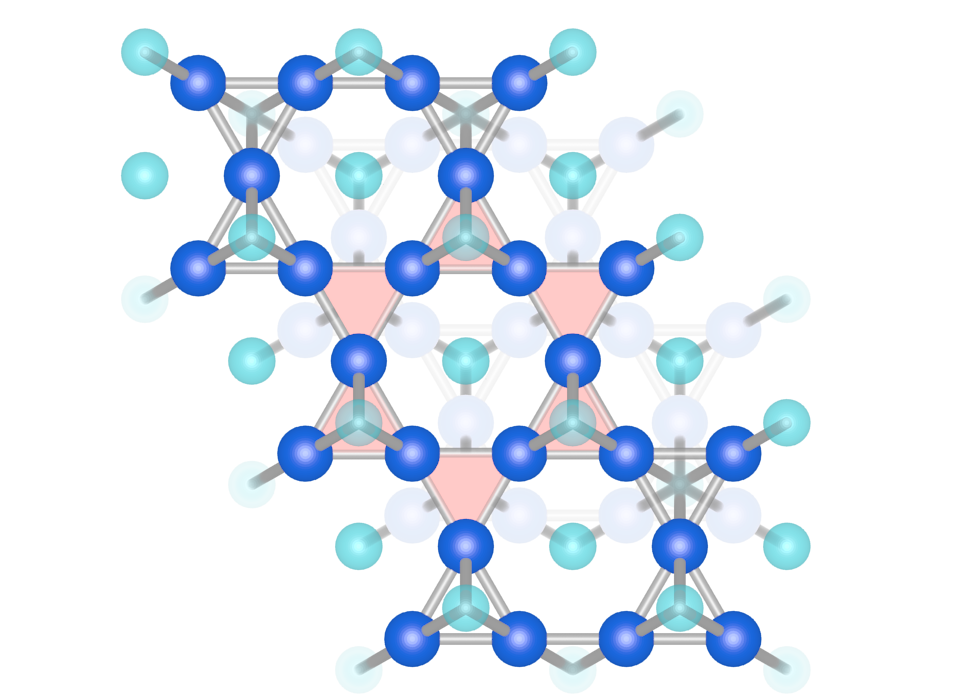}
			\caption{}
			\label{fig:tripod}
		\end{centering}
	\end{subfigure}
	
	\begin{subfigure}[t]{0.46\textwidth}
		\begin{centering}
			\includegraphics[scale=0.32]{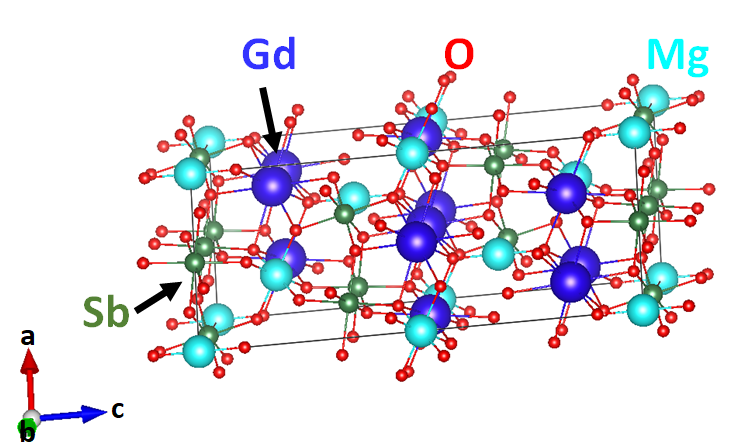}
			\caption{}
			\label{fig:unit_cell}
		\end{centering}
	\end{subfigure}
	\caption{(a) Side view of the principal Mg$^{2+}$ and Gd$^{3+}$ ions which constitute the pyrochlore lattice. (b) View along $c$-direction of Mg$^{2+}$ and Gd$^{3+}$ ions with the kagome plane as red triangles. (c) Complete unit cell of $\text{Mg}_{2}\text{Gd}_{3}\text{Sb}_{3}\text{O}_{14}$.}
	\label{fig:pyrochlore_vs_tripod}
\end{figure}

Recently, a new class of interesting compounds, called tripod kagome (TK) materials with the empirical formula $\text{Mg}_{2}\text{RE}_{3}\text{Sb}_{3}\text{O}_{14}$ (RE\,=\,rare earth), have been synthesized and characterized \cite{RE_tripod_1,RE_tripod_2}, among them the title compound $\text{Mg}_{2}\text{Gd}_{3}\text{Sb}_{3}\text{O}_{14}$. This compound  is an example of an effective 2D magnetic planar structure, with kagome lattice planes consisting of magnetic $\text{Gd}^{3+}$ ions with spin $S$\,=\,7/2 separated by planes of nonmagnetic $\text{Mg}^{2+}$ ions (see Fig. \ref{fig:pyrochlore_vs_tripod}). Thus, Mg$_2$Gd$_3$Sb$_3$O$_{14}$ appears to be a 2D-counterpart of, e.g., the well-known compound $\text{Gd}_{2}\text{Ti}_{2}\text{O}_{7}$ \cite{spin_dynamics_Gd2Ti2O7,Ramirez2002}, in which the $\text{Gd}^{3+}$  ions sit in a frustrated pyrochlore lattice, making it an inherent 3D magnet. The spin-orbit coupling in Gd$^{3+}$ ions, which possess a half-filled 4$f$-shell, vanishes up to first order. Thus, a splitting of the $(2S+1)-$degenerate spin multiplet of the ion due to the crystal field of the oxygen ligands is expected to be insignificant as compared to the relevant experimental magnetic energy scales of this material. While in $\text{Gd}_{2}\text{Ti}_{2}\text{O}_{7}$, phase transitions both depending on temperature as well as on an externally applied magnetic field have been observed \cite{Ramirez2002}, it is to be expected that in a 2D kagome lattice magnet incorporating ions without spin-orbit interaction (such as $\text{Gd}^{3+}$), no long-range antiferromagnetic (AFM) or ferromagnetic (FM) ordering above $T$\,=\,0\,K will occur due to the isotropy of the exchange interaction, if there are no additional anisotropic contributions \cite{Mermin1966} or interlayer couplings. However, Dun {\it et al.} did observe an AFM phase transition at $T_{\rm N} = 1.65$\,K for $\text{Mg}_{2}\text{Gd}_{3}\text{Sb}_{3}\text{O}_{14}$ \cite{RE_tripod_1}, suggesting that there must be some anisotropic interaction between the $\text{Gd}^{3+}$ spins or a residual interlayer coupling. Significant anisotropies might arise, for example, as a result of (classical) dipolar spin-spin interactions between the $\text{Gd}^{3+}$ ions within the kagome layer. These findings motivate a more detailed study of static and dynamic magnetic properties of this compound. In this respect electron spin resonance (ESR) spectroscopy is the method of choice, as it can provide insights into the type and the strength of the spin-spin interactions, in particular with respect to magnetic anisotropies.

Here, we present results of an ESR spectroscopic study of a polycrystalline sample (powder) of \MgGd\  in a broad frequency, magnetic field and temperature range complemented by measurements of the static magnetization $M(H,T)$. Our static magnetic susceptibility data $\chi(T) = M_{H}(T)/H$ reveal the presence of an AFM phase transition at 1.7\,K, similar to Ref.~\cite{RE_tripod_1}. We find that this transition is rapidly suppressed to temperatures below 0.5\,K by applying fields larger than $\mu_0H \sim  3$\,T. The analysis of the $M(H)$ dependences at low $T$ reveals an isotropic exchange coupling $|J|\approx 0.3$\,K (with energy $\epsilon_{\text{ex}} = JS^{2}\approx 3.5$\,K for classical spins). In our high-field ESR experiments we observed a remarkable evolution of the shape of the ESR response from a single Gaussian line to a broad, asymmetric line profile featuring a high-field shoulder which gradually develops for $T < 30$\,K. We analyzed in detail different mechanisms which could be responsible for this behavior, such as demagnetizing fields, single-ion anisotropy, and spin-spin correlations and discuss their possible relevance to the experimental observations.

The paper is organized as follows. In Sect.~\ref{sec: experimental} sample preparation and characterization as well as technical experimental details are described. Static magnetometry and high-field ESR results are presented in Sect.~\ref{sec: results} and discussed in Sect.~\ref{sec: discussion}. In Sect.~\ref{sec: conclusions} main conclusions are formulated. Details of the data analysis are explained in the appendix, Sect.~\ref{sec: appendix}.

\section{Synthesis, Sample Characterization and Experimental Details}\label{sec: experimental}
A white polycrystalline sample of {\MgGd} was prepared through a solid-state reaction using a stoichiometric mixture of dried Gd$_2$O$_3$ (99.9\,\%, Aldrich), MgO (99.95\,\%, Alfa Aesar) and Sb$_2$O$_5$ (99.9\,\%, Alfa Aesar). The mixture of the precursor compounds was homogenized by grinding with mortar and pestle, followed by a 24\,h sintering at 1000$^{\circ}$C. The sample was subsequently ground, pressed into pellets and fired at 1300$^{\circ}$C for 9~h. Phase purity of the product was assessed by powder X-ray diffraction (XRD), by using a STOE Stadi P powder diffractometer in transmission geometry with Cu K$_{\alpha 1}$ radiation. The diffractometer is equipped with a curved Ge (111) monochromator and a 6$^{\circ}$ linear position sensitive detector (DECTRIS MYTHEN 1\,K detector). 


\begin{figure*}[t]
	\includegraphics[scale=0.54]{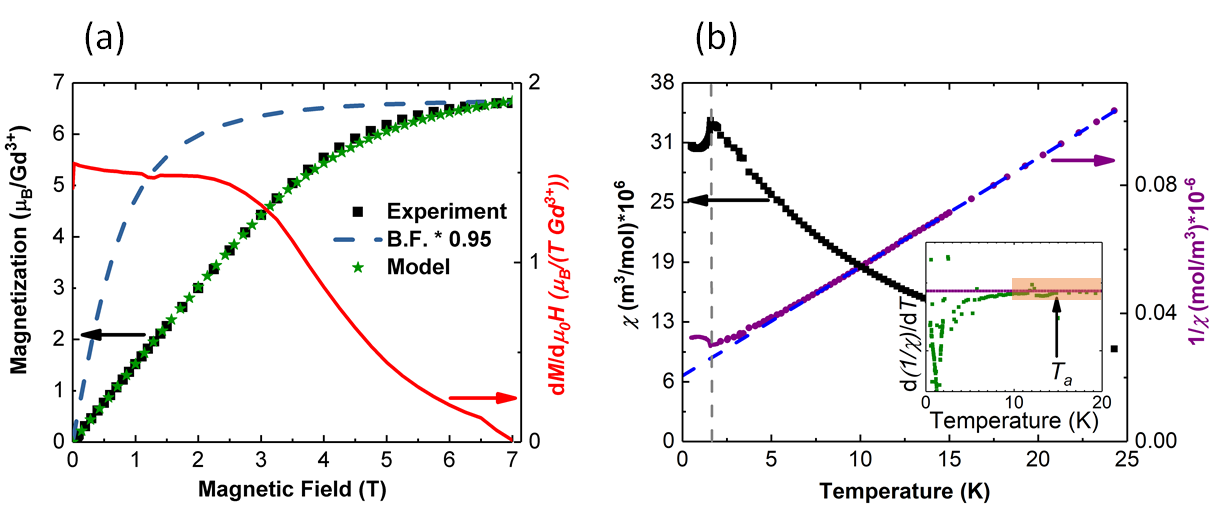}
	\caption{
(a) Field dependence of the magnetization at $T$\,=\,$2$\,K (black squares), its derivative $dM/dH$ (solid line), Brillouin function corresponding to non-interacting Gd$^{3+}$ ions with $J=S=7/2$ (dashed line, scaled to the observed saturation moment), and the modelled $M(H)$ dependence (green stars) which takes into account dipolar and exchange interactions between the Gd$^{3+}$ spins (for details see the text). (b) Temperature dependence of the static susceptibility $\chi (T)$ (black squares) and 1/$\chi (T)$ (purple dots) at a field of $\mu_0H=20$\,mT together with the Curie-Weiss fit (dashed line). The AFM transition at $T_{\rm N}$\,=\,1.7\,K is indicated by the vertical dashed line. The insert shows the derivative of 1/$\chi(T)$ (symbols). The horizontal solid line corresponds to the derivative of the high-temperature Curie-Weiss fit. The shaded rectangle centered around $T_{\rm a}\approx 15 \pm 5$\,K  indicates the region where the $\chi(T)$ dependence begins to gradually deviate from the Curie-Weiss law.}

	\label{fig:static_magnetism}
\end{figure*}
\noindent

\begin{figure}
	\begin{centering}
		\includegraphics[scale=0.33]{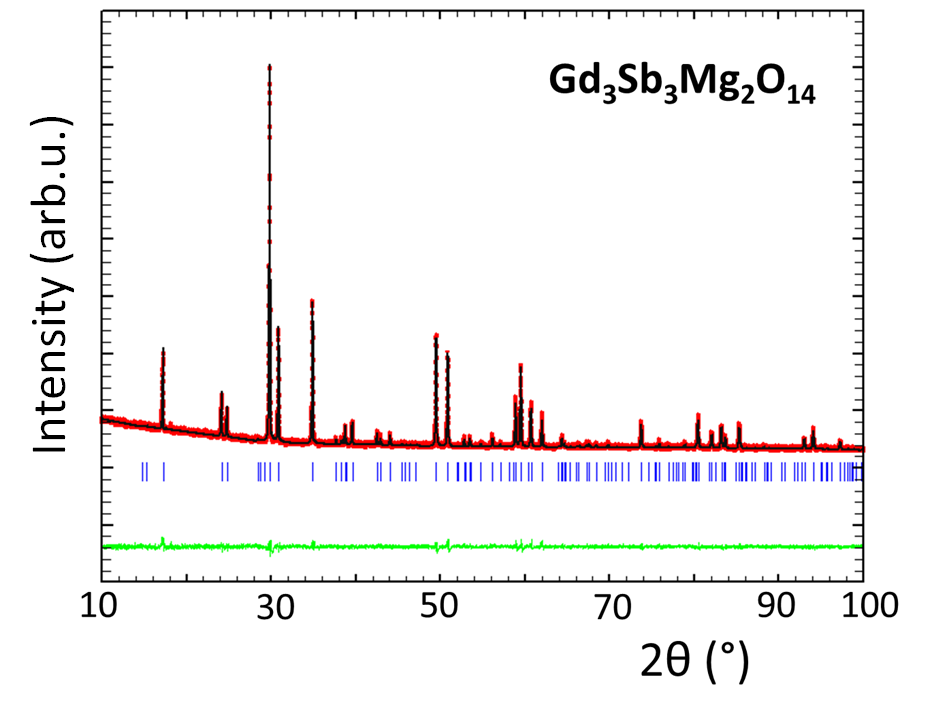}
		\caption{Calculated (black line) and observed (red dots) XRD patterns and the difference thereof (green line) of the Rietveld refinement for \MgGd ($\lambda$ = 1.5460\,\AA, Bragg R-factor: 5.61\,\%; Rf-factor = 7.17\,\%; Bragg R-factor = $\Sigma |\text{I}_{k-obs} - \text{I}_{k-calc}| / \Sigma|\text{I}_{k-obs}|$; Rf-factor = $[(N - P)/\Sigma \text{w}_{i} \text{y}^2_{i-obs}]^{1/2}$). The blue lines indicate the Bragg reflections. The inset shows an optical image of a typical white polycrystalline sample of {\MgGd}.}
		\label{fig:Rietveld}
	\end{centering}
\end{figure}

The single-phase nature of the material was confirmed by the match between the calculated and the experimental patterns (Bragg R-factor = 5.61, Rf-factor = 7.17) based on Rietveld analysis of the powder XRD data (see Fig. \ref{fig:Rietveld} in Sec.~\ref{appendix4}). \MgGd ~crystallizes in a rhombohedral structure with $R-3m$ space group in a hexagonal coordinate system, with cell parameters of $a$\,=\,7.3556(1)\,\AA, $c$\,=\,17.350(2)\,\AA. The crystallographic parameters according to our structural model are similar to those previously reported~\cite{RE_tripod_1,RE_tripod_2,structure_3}. For details regarding the Rietveld analysis and the structural parameters, the reader is referred to the Appendix, Sect.~\ref{appendix4}.

Magnetization measurements were carried out with Superconducting Quantum Interference Device (SQUID) magnetometers by Quantum Design (SQUID-VSM and MPMS-XL). The field dependence of the magnetization was recorded at $T=300$\,K and $T$\,=\,2\,K in a range of $\mu_0H\,=\,0-7$\,T, while the temperature dependence of the magnetization was recorded in zero-field cooled mode at $\mu_0 H=20$\,mT in a range of $T\,=\,0.5-300$\,K. In the low temperature range of $T\,=\,0.5-2$\,K, also the field-cooled mode was employed and $\chi(T)$ measured at $\mu_0 H$\,=\,$0.02-5$\,T. A commercial Helium-3 insert was used for measurements below 2\,K. 

Electron spin resonance (ESR) measurements were conducted using a Bruker EMX X-band spectrometer, which operates at a microwave frequency of 9.6\,GHz in magnetic fields up to 0.9\,T, within a temperature range of $T=4-295$\,K. For the ESR experiments at higher frequencies and fields, a vector network analyzer from Keysight Technologies (PNA-X) together with extensions from Virginia Diodes were employed, covering a frequency range of $70-330$\,GHz. In order to generate high magnetic fields up to 16\,T at various sample temperatures, a cryomagnet from Oxford Instruments was utilized. Temperature dependent spectra were recorded at temperatures in a range of $T=3-40$\,K. Before mounting the sample into the high-field ESR setup, the powder was diluted with epoxy to prevent residual interactions of the powder particles as well as an orientation of the powder particles along a preferred axis, in contrast to the magnetization measurements, where a loose powder sample was used. To carry out measurements in a broad frequency range, an oversized waveguide without a resonating cavity is used in the high-field ESR experiments. 
%
%
Possible distortions of the ESR lines due to the complex impedance of the quasi-optical path of the setup can be minimized by the vector network analyzer which measures both the amplitude and the phase of the transmitted signal, as well as by accounting for both absorptive and dispersive components in the analysis of the ESR lineshapes~\cite{lineshape_formulas,Kendall2003}.  


%
%


%

\section{Experimental Results}\label{sec: results}
\subsection{Magnetization and Susceptibility}\label{magnetization}
%
%

In Fig.~\ref{fig:static_magnetism}, the results of the magnetization $M(H)$ and the susceptibility $\chi (T)$ measurements are shown. The magnetization curve measured at $T$\,=\,2\,K as well as its derivative [see Fig.~\ref{fig:static_magnetism}(a)] exhibit a continuous, monotonic behavior with no signs of a phase transition, in close agreement with the earlier data in Ref.~\cite{RE_tripod_1}. At the highest applicable field of $\mu_0H=7$\,T, the moment per Gd$^{3+}$ ion amounts to $\mu \approx 6.6\,\mu_{\text{B}}$, which is about 94\% of the theoretical saturation moment $\mu_{\text{sat}} = gS\mu_{\text{B}} = 7.0\,\mu_{\text{B}}$ for a $g$-factor of $2.0$, $S = 7/2$, and Bohr magneton $\mu_{\text{B}}$. The $M(H)$ curve follows a linear behavior up to approximately
$\mu_{\text{0}}H$\,=\,3\,T, until saturation effects set in. This is in stark contrast to an $M(H)$ dependence of non-interacting paramagnetic ions with a large spin value $S = 7/2$, which saturates much faster at such a low temperature, indicating a significant influence of antiferromagnetic interactions at 2\,K (see Sec.~\ref{sec: Magnetization_Discussion} and Sec.~\ref{appendix1} for a model and details).

In Fig.~\ref{fig:static_magnetism}(b) the temperature dependent susceptibility $\chi(T)$ and its inverse 1/$\chi(T)$ at a field of $\mu_0H=20$\,mT are shown. A Curie Weiss fit from $70-300$\,K yields an effective magnetic moment of $\mu_{\text{eff}}=7.8\mu_{B}$ and a Weiss temperature of $\theta_{W}=-6 \pm 1$\,K, which agrees well with the literature data \cite{RE_tripod_1} and is consistent with an effective moment of free $\text{Gd}^{3+}$ ions ($^{8}\text{S}_{7/2}$ term), $\mu_{\text{eff}}=7.94\,\mu_{B}$. A magnetic phase transition at a temperature of $T_{\rm N}$\,=\,1.7\,K is discernible by a broad maximum in the susceptibility in agreement with the ac susceptibility and specific heat data in Ref. ~\cite{RE_tripod_1}. This value lies between $|\theta_{W}|$ (transition temperature expected for an unfrustrated 3D magnet) and 0\,K (i.e. no transition, expected for an unfrustrated isotropic 2D magnet).

\begin{figure}
	\includegraphics[scale=0.3]{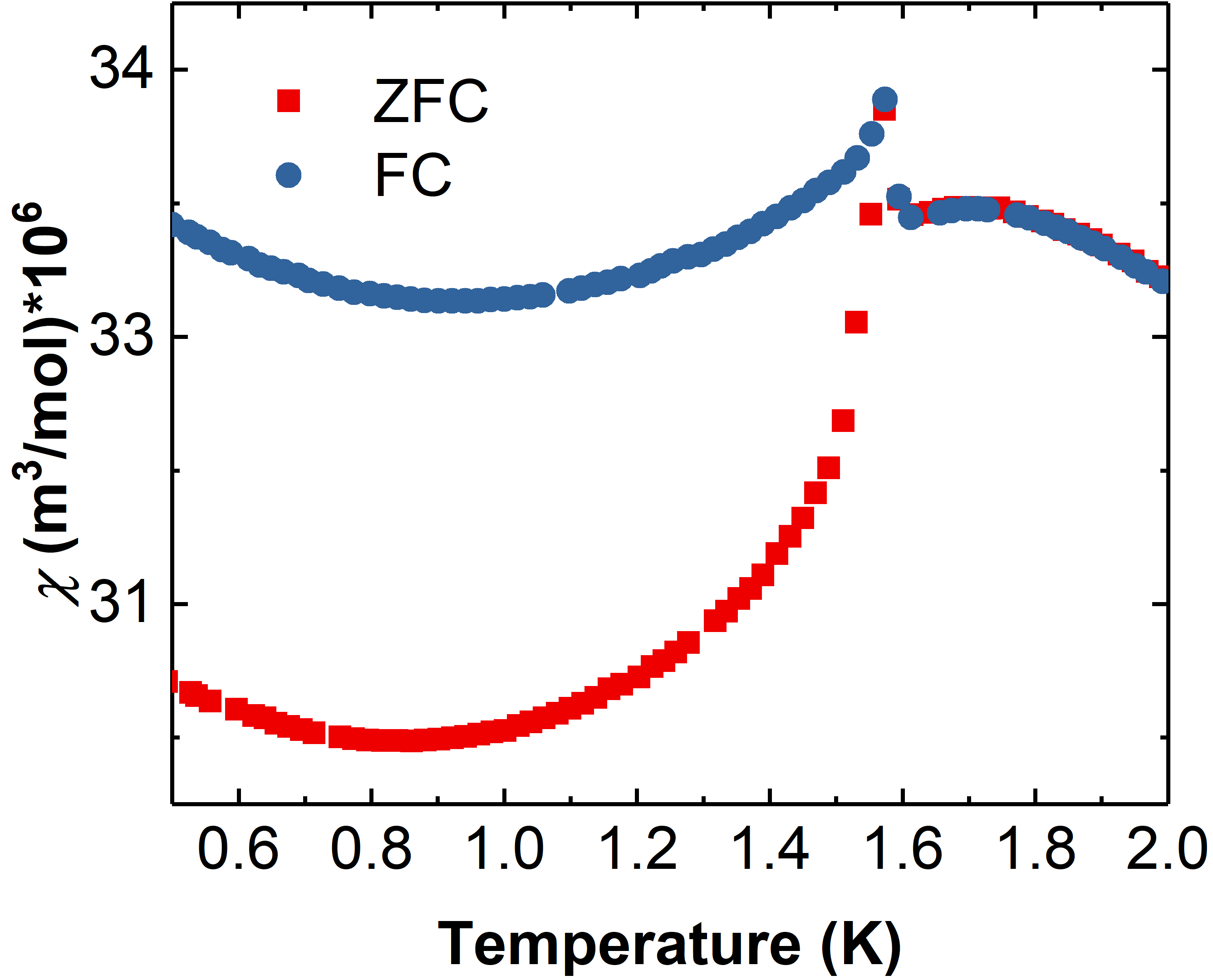}
	\caption{Zero field cooled (ZFC, red squares) vs. field cooled (FC, blue circles) susceptibility at low temperatures measured at a field $\mu_{0}H$\,=\,20\,mT.}
	\label{fig:ZFC_FC_Susceptibility}
\end{figure}

Besides the phase transition one finds a characteristic temperature region around a loosely defined temperature $T_{\text{a}} \approx 15 \pm 5$\,K, where the susceptibility starts to deviate gradually from the paramagnetic Curie-Weiss law, as can be seen in the 1/$\chi$ and derivative plots in Fig.~\ref{fig:static_magnetism}(b). Such a deviation typically indicates the entry of a correlated regime at temperatures $\lesssim 15$\,K (see Sec.~\ref{sec: Magnetization_Discussion} for a further discussion). 

\begin{figure}[H]
	\centering
	\includegraphics[scale=0.3]{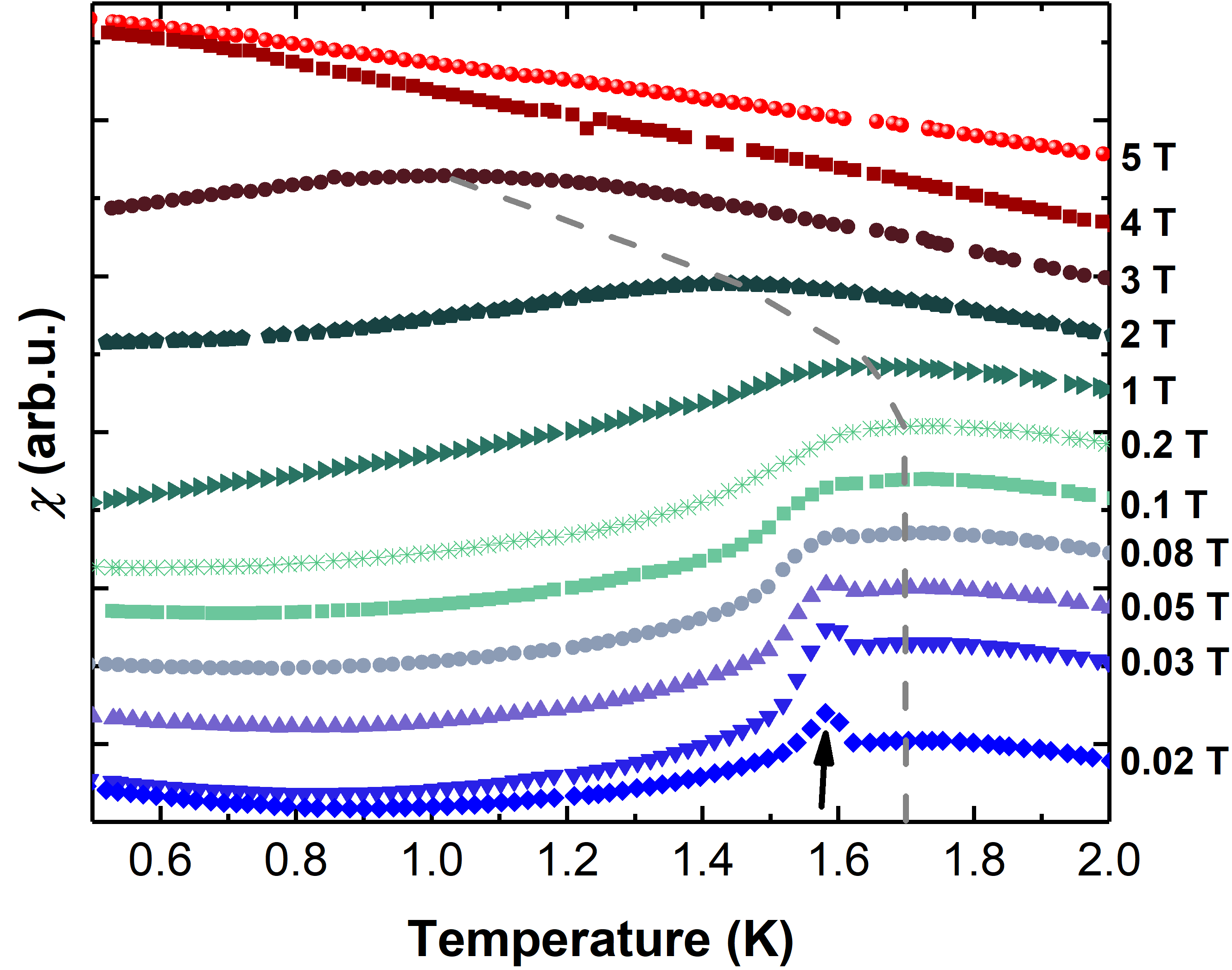}
	\caption{Temperature dependence of the static susceptibility $\chi(T)$ in the low-$T$ range at different applied magnetic fields. The curves are shifted with respect to each other in order to improve visibility. The grey dashed line marks the position of the broad maximum (named $T^{\rm max}$ in the text) of the $\chi(T)$ curves. The arrow shows the sharp peak on top of the broad maximum.}
	\label{fig:Field_Dep_Susceptibility}
\end{figure}

Interestingly, below $T_{\rm N}$ the $\chi(T)$ measured in the field cooled (FC) and zero field cooled (ZFC) modes at $\mu_{0}H$\,=\,20\,mT deviate from each other, with a stronger decrease in the ZFC curve (Fig.~\ref{fig:ZFC_FC_Susceptibility}). Furthermore, a sharp peak on top of the broad maximum is visible at this scale and the splitting between ZFC and FC magnetization ends up at the summit of this peak. The sharp peak becomes less prominent at higher magnetic fields and vanishes at $\mu_0 H \geq 0.1$\,T (Fig.~\ref{fig:Field_Dep_Susceptibility}). With regard to the previously proposed ground state~\cite{RE_tripod_1}, these are somewhat surprising features which give rise to several tentative scenarios of magnetic order, discussed in Sec.~\ref{sec: Susceptibility_Discussion}. 

\begin{figure*}
	\hspace*{1cm}\includegraphics[scale=0.47]{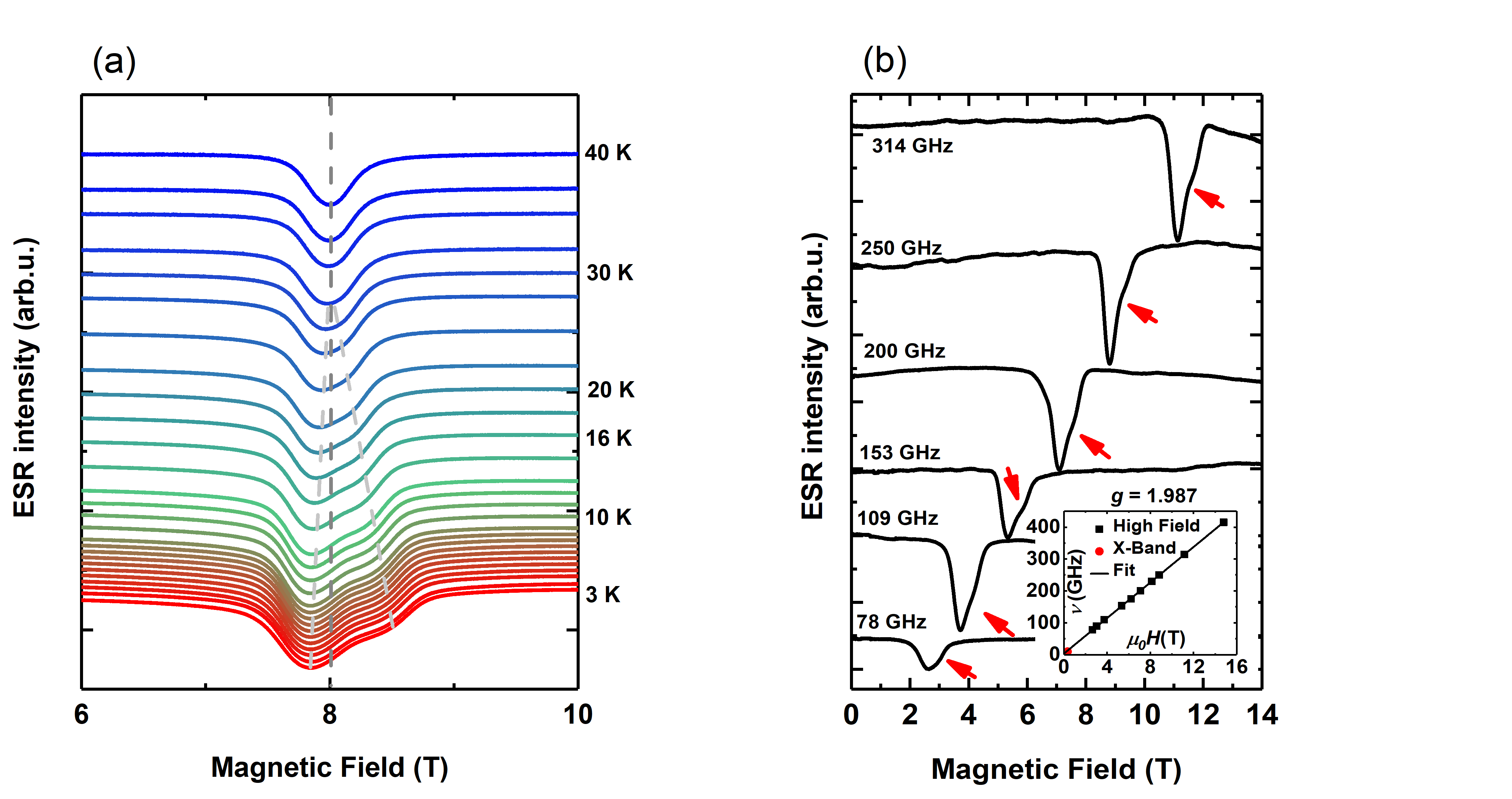}
	\caption{(a) Temperature dependence of the ESR spectrum of \MgGd\ at $\nu = 222$\,GHz. The vertical dashed line indicates the high-temperature resonance field. Note the development of the shoulder at the high-field side of the spectrum with lowering $T$. The grey dashed lines show the temperature trend of the main peak and the shoulder; (b) Frequency dependence of the ESR spectrum at $T=3$\,K. The arrows indicate the shoulder in the spectra. Inset: Symbols denote the microwave frequency $\nu$ vs. the position of the main peak $H_{\rm res}$. The solid line shows the fit of the data to the resonance condition $h\nu$\,=\,$g\mu_{\rm B}H_{\rm res}$ yielding the g-factor $g$\,=\,1.987.}
	\label{fig:Summary}
\end{figure*}

\subsection{High-field ESR}\label{ESR}
Typical high-field ESR spectra for \MgGd~are shown in Fig.~\ref{fig:Summary} for selected frequencies and temperatures. 

At high $T$, the spectrum consists of a single Gaussian-shaped line. A shoulder at the high-field side of the signal begins to develop below $\sim 30$\,K [Fig.~\ref{fig:Summary}(a)]. The main resonance peak $H_{\rm res}$ shifts by an amount of $\approx~-0.16$\,T at $\nu = 222$\,GHz and $T=3$\,K with respect to its position at $T=40$\,K ($\approx 8$\,T) whereas the shoulder shifts by $\approx + 0.5$\,T to the high-field side. Note that this change of the spectral shape is continuous, showing no signs of a magnetic phase transition, which, according to the susceptibility data, takes place at a much smaller temperature and only at fields smaller than $\sim 4$\,T (Sect.~\ref{magnetization}). The frequency dependence of the ESR spectrum shows that the shoulder on the right side of the main peak can be clearly discerned at frequencies $\geq 80$\,GHz [Fig.~\ref{fig:Summary}(b)]. In the inset to Fig.~\ref{fig:Summary}(b) the frequency dependence of the main peak position $\nu(H_{\rm res})$ at $T$\,=\,$3$\,K, which was determined by picking the minima of the spectra, is plotted together with a linear fit according to the resonance condition $h\nu$\,=\,$g\mu_{\text{B}}H_{\rm res}$. 
The fit yields the g-factor $g\,=\,1.987$, which matches with the free-ion value for Gd$^{3+}$, as expected \cite{Abragam_Bleaney_1970}. This value is validated by a fit at $T\,=\,40$\,K, i.e. in the purely paramagnetic regime, which gives $g\,=\,1.983$ (see Fig.~\ref{fig:Xband_gfactor40K}(b) in the Appendix). The linear $\nu$ vs. $H_{\rm res}$ relationship shown in the inset to Fig.~\ref{fig:Summary}(b) suggests a gapless nature of the ESR excitation. Indeed, ESR measurements at a small X-band microwave frequency of $\nu\,=\,9.6$\,GHz (see Appendix~\ref{appendix5}) evidence that the excitation gap at $T\,=\,4$\,K, if present at all, should be less than $\nu\,=\,10$\,GHz, corresponding to a field of $\mu_{\text{B}}H\,=\,0.34$\,T with a g-factor of 2.
The most remarkable observation in the presented high-field ESR experiments is the occurrence of the shoulder feature in the ESR spectrum of \MgGd\ continuously developing over a broad temperature range below 30\,K. The possible reasons for such a peculiar evolution of the ESR spectrum and its relation to the physics of \MgGd\ will be discussed in detail in the subsequent Sect.~\ref{sec: ESR_Discussion}. 


\section{Data Analysis and Discussion}\label{sec: discussion}

\subsection{Magnetization $M(H)$}\label{sec: Magnetization_Discussion}
The experimental $M(H)$ curve shown in Fig.~\ref{fig:static_magnetism}(a) differs significantly from a Brillouin function $B_{J}(H,T)$ for a free Gd$^{3+}$ ion at $T = 2$\,K plotted on the same figure. Such a deviation of the experimental dependence from the behavior expected for free ions is usually an indication of significant magnetic interactions of dipolar and possibly of exchange nature which generate effective internal fields in the material. The following analysis along this scenario is aimed to estimate the strength of these interactions in \MgGd\ responsible for the shape of the magnetization curve. 

In our approach we include dipolar and exchange interactions by converting them into effective fields acting on the spin sites. The effective total field $\vec{H}_{\text{tot}}$ in the Brillouin function then reads $\vec{H}_{\text{tot}} = \vec{H}_{\text{ext}} + \vec{H}_{\text{DD}} + \vec{H}_{\text{J}}$ with the external, dipolar and exchange fields, respectively.

The effective (static) dipolar and exchange fields arise from average polarization moments of the spins in an external magnetic field and are as such proportional to the magnetization and thus dependent on the external field and the temperature. The analysis shows that the effect of the exchange interaction is substantial for the reproduction of the experimental $M(H)$ curves (see Appendix, Sect.~\ref{appendix1}, for details). The mean exchange field takes the form $\vec{H}_{\text{J}}$=$a_{\text{ex}}n_{\text{NN}}M\vec{b}$, with $a_{\text{ex}}$ the exchange parameter, $n_{\text{NN}}$ the number of nearest neighbors for each ion site, $M$ the magnetization and $\vec{b}$ the unit vector in the direction of the external field. The resulting modelled magnetization curve with the optimized exchange parameter $a_{\text{ex}}$\,=\,-0.11\,T is shown in Fig.~\ref{fig:static_magnetism}(a).

From this model, the dipolar field strength for a powder averaged polarization of the spins can be calculated and compared to the internal field strength according to the internal field model for the ESR data (see Sect.~\ref{ESR}). The dipolar field amounts to $\mu_{\text{0}}H_{\text{DD}}$\,=\,0.67\,T (or in terms of dipolar energy, $E_{\text{DD}} $\,=\,3.15\,K) for $T$\,=\,3\,K and $\mu_{\text{0}}H$\,=\,8\,T, the resonance field at $\nu$\,=222\,GHz. There is a discrepancy between this value and the dipolar energy value found from ESR measurements (2.35\,K), which is discussed in the end of Sect.~\ref{ESR}.

Translating the value of the exchange field $a_{\text{ex}}$ into an isotropic exchange constant $J$ for an interaction of the form $J\vec{S}_{\text{i}}\vec{S}_{\text{j}}$ gives $J \approx $ 0.3\,K (AFM interaction) and an exchange interaction energy for two spins of $\epsilon_{\text{ex}}~\approx $~3.5\,K (see Appendix, Sect.~\ref{appendix1} for further explanation). In a kagome system, each spin has four nearest neighbors, so the total exchange interaction energy on one spin site is of the order $E_{\text{ex}}$\,=\,$n_{\text{NN}}\epsilon_{\text{ex}}$\,=\,14\,K. This is much larger than the dipolar interaction energy. Additional $M(H)$-plots demonstrating a very good agreement between experimental and model data at different temperatures both below and above the phase transition are shown in Appendix~\ref{appendix1} (Fig.~\ref{fig:Temp_Dep_M_over_H}). Those results show that the model provides a consistent explanation of the experimental $M(H)$ dependence in a large temperature range of at least $T = 0.5 - 30$\,K, and that a mean field description of the $M(H)$ curves with the above estimated $T$-independent exchange parameter $a_{\text{ex}}$ is adequate even at very low temperatures and even in the ordered phase.

\subsection{Susceptibility $\chi(T)$}\label{sec: Susceptibility_Discussion}
The susceptibility curve depicted in Fig.~\ref{fig:static_magnetism}(b) shows several characteristic features, which are discussed in this section. First of all, $T_{\text{N}} = 1.7$\,K is only approximately 1/4 of the Curie-Weiss temperature $|\theta_{\text{CW}}|$. This rather small value of $T_{\text{N}}$ may indicate both a pronounced magnetic low-dimensionality of the well-separated Kagome layers and the resulting 2D physics, and inherent geometrical frustration of the Gd$^{3+}$ spin network. Interestingly, with the value of the isotropic exchange constant $J\approx 0.3$\,K estimated in the previous Sect.~\ref{sec: Magnetization_Discussion} one obtains a standard mean-field estimate of $\theta_{\text{CW}}$ as $|\theta_{\text{CW}}| = n_{\rm NN}JS(S+1)/3 \approx 6$\,K. Such a close agreement with experiment further supports the validity of the mean-field model for the description of the static magnetism of \MgGd.

Secondly, the $\chi(T)$ curve starts deviating gradually from the high-temperature Curie-Weiss dependence around a  temperature of $T_{\text{a}} \approx 15 \pm 5$\,K [Fig.~\ref{fig:static_magnetism}(b), inset]. Such a deviation is typically attributed to an onset of spin-spin correlations by approaching a magnetic phase transition, or a change in a temperature dependent effective magnetic moment. The latter can be excluded as an explanation for the susceptibility behavior, since the thermal depopulation of the fine structure-split spin levels of the Gd$^{3+}$ ion, which could possibly reduce the magnetic moment, occurs at a much lower temperature (see Sect.~\ref{sec: ESR_Discussion}) and the $g$-factor is temperature independent. $T_{\text{a}}$ turns out to be of the same order as $JS^{2}n_{\text{NN}} \approx$ 14\,K. Up to this point, these findings suggest that this system with an effective $S = 7/2$ can be viewed as a classical 3D Heisenberg paramagnet above a characteristic temperature $T_{\mathrm{a}}\geq 15$\,K, as far as static properties are concerned, but gradually enters a correlated regime at smaller temperature.

Thirdly, the FC curve differs from the ZFC below $T_{\text{N}}$, with a stronger decrease in the ZFC curve (Fig.~\ref{fig:ZFC_FC_Susceptibility}). Previous specific heat measurements suggest a long-range ordered ground state, and previous ac susceptibility measurements have shown no frequency dependence, excluding the possibility of a glassy state~\cite{RE_tripod_1}. However, the 120$^\circ$ phase with all spins lying in the $ab$-plane as the proposed long-range magnetic order in Ref.~\cite{RE_tripod_1} is not expected to give a splitting between ZFC and FC magnetization. Instead, one possibility could be a slight canting of the spins towards the $c$-axis resulting in a small ferromagnetic component. There are two possible canting directions along the $c$-axis, one of which would then get selected in FC mode, yielding a higher magnetization. Alternatively, a collinear AFM structure with moments pointing in the kagome plane may appear. In such a collinear AFM structure there would be three possible moment directions according to the rhombohedral symmetry of the lattice. Again, in FC mode the moment direction in each powder particle would be selected by the external magnetic field among the three possibilities, leading to a higher magnetization.

Finally, the low-temperature behavior of the susceptibility suggests a more complex phase transition. The occurrence of the sharp peak itself is not expected for a transition to an AFM ordered state of an ensemble of Heisenberg spins, but might be a peculiarity of  a 2D-Ising-like transition proposed for~\MgGd\ in Ref.~\cite{RE_tripod_1}. Its observation hints at another possible energy contribution in addition to the dipolar and isotropic exchange interactions considered here. A measurement of the susceptibility in the vicinity of the phase transition at different applied magnetic fields reveals that this sharp peak is absent for $\mu_0H\geq 0.1$\,T, and only the broad maximum $T^{\rm max}$ (where $d\chi(T)/dT = 0$) remains (Fig.~\ref{fig:Field_Dep_Susceptibility}). The latter shifts towards lower temperatures at higher fields, starting from $T^{\rm max} = 1.72$\,K at $\mu_0H = 0.02$\,T, then moving to 1.64\,K and 1\,K at 1\,T and 3\,T, respectively. Interestingly, a similar behavior (a peak structure on top of a broad transition) was observed in the compound Ba$_3$CoNb$_2$O$_9$~\cite{Lee2014}, possessing a layered triangular lattice of Co$^{2+}$ ions. This can be contrasted to Ba$_8$CoNb$_6$O$_{24}$, where the magnetic layers are very far apart and only a broad transition is left~\cite{Rawl2017}. However, in those cases, the transitions are similarly reflected in the magnetic specific heat, whereas in \MgGd, only the sharp transition could be observed~\cite{RE_tripod_1}.

\subsection{Electron Spin Resonance}\label{sec: ESR_Discussion}

The remarkable development of the asymmetry of the high-field ESR signal of \MgGd\ at low temperatures (Fig.~\ref{fig:Summary}) deserves a careful analysis with regard to the origin of this peculiar lineshape. 

The trivial reason for such kind of asymmetry arising from the powder averaging of an ESR signal with an anisotropic $g$-factor \cite{Poole1997} can be safely excluded since the $g$-factor of the spin-only Gd$^{3+}$ ion is practically isotropic. Moreover, this kind of anisotropy would result in a temperature independent effect as the spectroscopic $g$-tensor usually does not depend on temperature. Therefore, on needs to identify a relevant, temperature dependent {\it anisotropic} mechanism which could consistently explain the experimental observations.

Essentially, one can find out three major effects from which a significant contribution is conceivable: 

\noindent
(i) Due to a large magnetization of the gadolinium ions at high fields ($\mu_0H$\,=\,8\,T) and low temperatures, demagnetization effects may play a role. Depending on the shape of the powder particles, they may have an anisotropic contribution. 

\noindent
(ii) There may be a non-negligible influence of the crystal field acting on the Gd$^{3+}$ ions and leading to a significant single-ion anisotropy with strength $D$.

\noindent
(iii) Due to a slowing of the spin dynamics at low temperatures, anisotropic interactions between the Gd$^{3+}$ spins may become more prominent and result in enhanced anisotropic correlations between the spins. As an effect, quasi-static local anisotropic fields start to develop, which is similar in nature to the demagnetization. Such enhanced correlations in the paramagnetic state may be a clue to a possible frustration in the system. 

In the following, these three scenarios will be discussed in detail. 

\subsubsection{Demagnetization effects}\label{scenario:demag}

In order to treat possible demagnetization effects, the resonance condition $h\nu = g\mu_{\text{B}}H$ should be extended in such a way that $H$ includes both the externally applied magnetic field $H_{\text{0}}$ and demagnetizing fields $N_{\text{i}}M_{\text{i}}$, with $i$ = x,y,z and respective demagnetizing factors $\sum_i N_{\text{i}} = 1$.
Asymmetric powder-averaged lineshapes can be reproduced only when the geometry of the  particles leads to a strong anisotropy of the demagnetization contributions with a significant difference between the $N_{\text{i}}$. As a model, the free-energy approach discussed, e.g., in Refs.~\cite{Baselgia1988,Farle1998}, where the resonance fields are derived for arbitrary demagnetizing factors and external field directions, was used (see also, e.g., Refs.~\cite{Zeisner2019a,Zeisner2019b}, where this model has been successfully applied to high-field ESR data, with shape anisotropy taken into account). The results are shown in Fig.~\ref{fig:DemagnetizationResults} in the Appendix, Sec.~\ref{appendix2}. A uniaxial shape anisotropy (thin plates, $N_{\text{x}} = N_{\text{y}}$) was considered with both Lorentzian and Gaussian lines as basic lineshapes and $N_{\text{z}}$ ranging between $0.7 - 1.0$. $N_{\text{z}} = 1$ corresponds to an infinite plate and $N_{\text{z}} \sim 0.8$ corresponds to the actual shape of the powder particles in the studied samples with a finite maximal aspect ratio (thickness)\,:\,(lateral dimensions)\,$\approx 1 : 10$ \cite{Aharoni1998}, as revealed by optical microscopy. Possible mixing and a texture parameter for a preferred orientation of the powder particles were included in the model. For low temperatures and frequencies, good fit parameter sets could be found (Fig.~\ref{fig:DemagnetizationResults}(a), with $\nu\,=\,109$\,GHz, $T$\,=\,3\,K and $N_{\text{z}}$\,=\,0.9). However, for higher frequencies, $N_{\text{z}}$ needs to be reduced in order for the demagnetization effects not to become too strong (Fig.~\ref{fig:DemagnetizationResults} (b,c), with $\nu\,=\,176$\,GHz and optimal $N_{\text{z}}$\,=\,0.7, and $\nu\,=\,222$\,GHz and $N_{\text{z}}$\,=\,0.85). At higher temperatures, there is also the tendency for the optimal $N_{\text{z}}$ to become smaller [Fig.~\ref{fig:DemagnetizationResults}(a,d)], and the lineshape can be reproduced less good. On the other hand, reducing the $N_{\text{z}}$ value for low frequencies and temperatures makes the modelled shoulder feature become too weak [Fig.~\ref{fig:DemagnetizationResults}(a)]. Thus, demagnetization effects may have a significant impact on the lineshape in particular at low temperatures where the magnetization of the sample is large, but cannot account consistently for the evolution of the high-field ESR spectra in the whole temperature and frequency ranges covered by our experiments, as a variation of $N_{\text{z}}$ which was necessary for good fits is unphysical. 

\subsubsection{Single-ion anisotropy of the Gd$^{3+}$ ion}\label{scenario:SIA}
The single-ion anisotropy arises due to the splitting of the ground-state spin multiplet of a paramagnetic ion with $S>1/2$ by the crystal field. In the simplest case of a uniaxial symmetry the respective Hamiltonian reads $\hat{\cal H}_{\rm CF} = DS_{\rm z}^2$, where the constant $D$ parameterizes the strength of the single-ion anisotropy. The eight-fold degenerate spin multiplet of the Gd$^{3+}$ ($S = 7/2,\: L= 0$) would be then split into four doublets. Such a splitting gives rise to a fine-structure of the Gd$^{3+}$ ESR signal with seven lines extending over the field range $12Dk_{\rm B}/g\mu_{\rm B}$ \cite{Abragam_Bleaney_1970}. Note that, although in the case of Gd$^{3+}$ ions, the single-ion anisotropy is a second-order effect, an anisotropy arising from the crystal field in Gd$^{3+}$ may actually be quite sizeable in some cases, as, e.g., was shown in Ref.~\cite{Glazkov2005}. There, the ESR spectrum of non-interacting Gd$^{3+}$ ions diluted in the (Y$_{\rm 1-x}$Gd$_{\rm x}$)$_2$Ti$_2$O$_7$ crystal with the pyrochlore structure demonstrates the splitting of spectral components of about 2\,T, which corresponds to $|D|\approx 0.25$\,K. The isotropic exchange interaction may narrow the fine-structure split ESR spectrum into a single Lorentzian-shaped line. As argued in Ref.~\cite{Sosin2006}, such a mechanism, known as the exchange narrowing effect \cite{vanVleck1948,Anderson1953}, is responsible for the narrowing of the ESR signal into a Lorentzian line at high temperatures in the respective concentrated compound Gd$_2$Ti$_2$O$_7$. The exchange narrowing ceases at low temperatures due to the increase of the short-range spin correlations and redistribution of spin level populations, yielding a broadening and a shift of the ESR signal of a single crystal of Gd$_2$Ti$_2$O$_7$ \cite{Sosin2006}. In the case of the polycrystalline sample of \MgGd, the powder averaging effect of this broadening and shift may, in principle, give rise to the observed asymmetric transformation of the ESR lineshape (Fig.~\ref{fig:Summary}).

However, some important observations contradict this scenario. At temperatures $T\geq 30$\,K, i.e., in the purely paramagnetic regime, the central part of the ESR signal of \MgGd\  is a symmetric Gaussian-shaped line [Fig.~\ref{fig:Summary}(a)]. This is in conflict with the exchange narrowing mechanism which yields  a purely Lorentzian lineshape \cite{vanVleck1948,Anderson1953}. Also, one can straightforwardly calculate the second moment of the Gaussian line \cite{Low1960} taking the Gd--Gd distances from the crystal structure which yields the dipolar contribution to the linewidth of $\sim 0.25$\,T. This corresponds well with the width of the measured high-temperature ESR signal of \MgGd\ suggesting that the extent of the possible fine structure should be significantly smaller than the dipolar width.

If to assume that the broadening of the low-$T$ ESR signal of \MgGd\ of about $\delta H \sim 1$\,T at high fields (Fig.~\ref{fig:Summary})  were entirely due to the fine-structure splitting of the ESR spectrum extending over $12Dk_{\rm B}/g\mu_{\rm B}$, then with $\delta H \sim 12Dk_{\rm B}/g\mu_{\rm B} \sim 1$\,T one obtains $D \approx 0.11$\,K.
The estimate of the exchange constant from the analysis of the magnetization, $J$\,=\,0.3\,K (corresponding to 0.22\,T in field units) implies that the exchange interaction is too small to be able to narrow this low-$T$ signal into a single line at high temperatures, as, for that, $J \gg D$ (i.e., $J \sim 12D \approx 1.3$\,K) would be required~\cite{Anderson1953}. This makes exchange narrowing effects as the cause of the temperature dependence of the lineshape unlikely. 

\noindent

Finally, in order to investigate spin-level population effects \cite{Sosin2006} on the ESR spectrum of \MgGd, a numerical analysis of the ESR lineshape according to a model with a uniaxial single-ion anisotropy~\cite{Abragam_Bleaney_1970} was performed. Only via an unphysically large variation of $|D|$ between $0.06$\,K to 0.105\,K good fittings could be achieved over a temperature range of $3 - 30$\,K. 

\subsubsection{Spin-spin correlation effects}\label{scenario:corellations}
Having discussed the possible effects of demagnetizing fields and of the single-ion anisotropy in the above Sects.~\ref{scenario:demag} and \ref{scenario:SIA}, respectively, 
we finally consider spin-spin correlations. The Gaussian lineshape of the single ESR lines of the individual powder particles even in the paramagnetic regime at high temperatures hints at a random distribution of local internal fields \cite{Abragam_Bleaney_1970}. 
Such internal fields may develop due to a slowing down of the spin dynamics with a decrease of the temperature and thus growing spatial short-range correlations $<S_i^{\alpha}S_j^{\beta}>$ between the Gd$^{\text{3+}}$ spins. The spin-spin correlations may have a twofold effect: On one hand, due to the randomness of the local fields, the single lines in the powder spectrum are inhomogeneously broadened, such that they have a Gaussian lineshape. On the other hand, by application of an external field, those correlations may behave anisotropically, which may lead to a shoulder in the powder averaged spectrum. A rough model for such an ansotropy is given by including an internal field $\vec{H}_{\text{int}}$ into the resonance condition

\begin{equation}
 g\mu_{\text{B}}(||\vec{H}_{\text{int}} + \vec{H}_{\text{ext}}||) = h\nu, 
\label{rescond}
\end{equation}
\noindent
resulting in a total field 
 \begin{equation}
 H_{\text{tot}} = (H_{\text{int}}^2 + H_{\text{ext}}^2 - 2H_{\text{int}}H_{\text{ext}}\,\text{cos}\theta)^{1/2}.
 \label{totfield} 
 \end{equation}
\noindent
with $H_{\text{int}} || $z without loss of generality (see Appendix, Sect.~\ref{appendix2}, for details). Exemplary powder averaged fits (Fig.~\ref{fig:Exemplary_Fitting_Plots} in the Appendix) show that, even with this simplified model, the lineshapes of the high-field spectra can be reproduced, meaning that anisotropic spin-spin correlations may be an additional contribution to the asymmetry of these spectra. One possible candidate for such anisotropic correlations is the dipolar interaction between the Gd$^{3+}$ spins.

\section{Conclusions}\label{sec: conclusions}

In summary, we have studied the magnetic properties of powder samples of the tripod-kagome antiferromagnet \MgGd\ by static magnetization and high-field ESR measurements. We observed a splitting between ZFC and FC magnetization in the magnetically ordered state below $T = 1.6$\,K, which gives some constraints on the possible nature of the long range magnetic order in \MgGd. Neutron diffraction experiments may help to elucidate the nature of this magnetic order and the specific spin structure. We found that this AFM order is very fragile and can be suppressed down to temperatures  $< 0.5$\,K by the application of magnetic fields larger than 3\,T, however leaving the system in a highly correlated state above $T_{\text{N}}$. The analysis of the static magnetic data has enabled to estimate the strength of the isotropic AFM exchange interaction between the Gd$^{3+}$ spins $S = 7/2$ of about $|J| \approx$ 0.3\,K (or exchange energy $E_{\text{ex}} \approx 14$\,K, considering all nearest neighbors). 

Importantly, our results show that at low temperatures, the understanding of the specific shape of the high-field ESR powder spectra of \MgGd\ requires a rather complex analysis, since different effects may become relevant. In this regard, we came to the conclusion that the ESR lineshape may be in part influenced by demagnetization effects due to the large magnetization of the sample, while correlations effects with possible dipolar origin may also play a role.
In the latter scenario, the growth of the spin-spin correlations would give rise to an extended temperature regime of \MgGd\ below 30\,K with pronounced low-energy spin fluctuations far above the ordering temperature. In any case, the fact that the onset of the asymmetric lineshape and line broadening are observed at such elevated temperatures ($T \approx 30$\,K) suggests significant geometrical frustration of magnetic interactions in the two-dimensional kagome spin lattice of Mg$_2$Gd$_3$Sb$_3$O$_{14}$. Thus, \MgGd\ can be considered as an almost ideal realization of the two-dimensional network of isotropic Heisenberg spins $S=7/2$ on a strongly frustrated kagome lattice.

\section{Acknowledgments}
The authors acknowledge technical assistance of \"Ozg\"ul Karatas in X-band ESR experiments.
This work has been supported by the Deutsche Forschungsgemeinschaft (DFG) within the Collaborative Research Center SFB 1143 "Correlated Magnetism - From Frustration to Topology" (Project No. 247310070), and the Würzburg-Dresden Cluster of Excellence "ct.qmat - Complexity and Topology in Quantum Matter" (EXC 2147, Project No. 390858490), and through Grants STU 695/1-1 (M.S.) and WU 595/3-3 (S.W.).


\bibliography{literature_tripod_Kagome}

\clearpage

\section{Appendix}\label{sec: appendix}


\subsection{Analysis of the tripod kagome (TK) structure}\label{appendix4}

\begin{table*}[t]
	\centering
	\caption{Structure parameters of \MgGd.}
	\begin{tabular*}{0.9\textwidth}{@{\extracolsep{\fill} } ccccccc}
		\textbf{Label} & \parbox[t]{1.7cm}{\textbf{Wyckoff \\ position}} & \textbf{x} & \textbf{y} & \textbf{z} & \textbf{Occupancy} & \textbf{B$_{\text{iso}}$ (\r{A})} \\
		\hline
		\\
		Mg(1) & 3$a$ (A) & 0 & 0 & 0 & 1 & 0.012(8) \\
		Mg(2) & 3$b$ (B) & 0 & 0 & 0.5 & 1 & 0.019(5) \\
		Gd & 9$e$ (A') & 0.5 & 0 & 0.5 & 1 & 0.018(9) \\
		Sb & 9$d$ (B') & 0.5 & 0 & 0 & 1 & 0.007(2) \\
		O(1) & 6$c$ & 0 & 0 & 0.3888(2) & 1 & 0.017(4) \\
		O(2) & 18$h$ & 0.5271(9) & 0.4729(9) & 0.8879(7) & 1 & 0.057(8) \\
		O(3) & 18$h$ & 0.4652(8) & 0.5348(8) & 0.3529(6) & 1 & 0.065(5)
	\end{tabular*}
	\label{table:Structure_Parameters}
\end{table*}
The structural formula of the TK compounds with a pyrochlore related structure can be written as AA'$_3$BB'$_3$O$_{14}$, in which A and A' are 8-fold and B and B' are 6-fold coordinated (see Fig.~\ref{fig:pyrochlore_vs_tripod}). In \MgGd, the sites A and B are fully occupied by Mg, while A' is fully occupied by Gd and B' by Sb. The ordering of the cations is driven by the ionic radius difference between the Mg$^{\text{2+}}$ and Gd$^{\text{3+}}$, which is larger than in the structurally similar compound Ca$_2$La$_3$Sb$_3$O$_{14}$. Therefore, in the Mg-based structure, the cationic sites are fully ordered, in contrast to the situation of the Ca-based material, Ca$_2$La$_3$Sb$_3$O$_{14}$, where the disorder between Ca$^{\text{2+}}$ and La$^{\text{3+}}$ ions was considerable because of their similar ionic radii \cite{Piccinelli1}.

The powder XRD data were analyzed with the Rietveld method using the FULLPROF in the WinPlotR program package~\cite{structure_1,structure_2}.  The peak shape was assumed to be a pseudo-Voigt function and the refinement included the following aspects: (i) the background, which was fitted using linear interpolation between 28 selected points; (ii) the scale factors; (iii) the global instrumental parameters (zero-point 2$\theta$ shift and systematic shifts, depending on transparency and off-centering of the sample); (iv) the lattice parameters, asymmetries and the overall temperature factor; (v) the profile parameters (Caglioti half-width parameters of the pseudo-Voigt function). The texture correction was included using the March-Dollase function.  The structure parameters are listed in Table~\ref{table:Structure_Parameters}. The positions of the atoms within a unit cell are in a good agreement with the values which have been previously reported \cite{RE_tripod_1,RE_tripod_2,structure_3}.

\noindent

\subsection{Details of the analysis of the $M(H)$ dependence}\label{appendix1}

\begin{figure}
		\begin{centering}
			\includegraphics[scale=0.3]{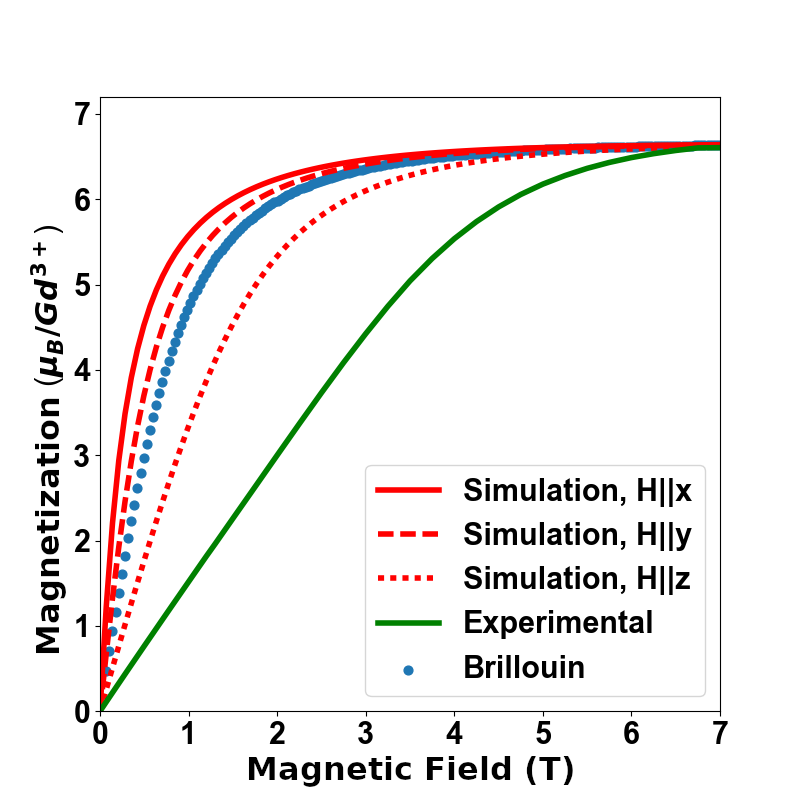}
	\caption{Simulated field dependence of the magnetization with only dipolar interactions in $x,y,z$-directions together with the experimental $M(H)$ curve and the Brillouin function of non-interacting spins $S=7/2$, at $T$\,=\,2\,K.}
		\label{fig:Mag_Effective_Field_Only_Dipolar}
		\end{centering}
	
\end{figure}

The field dependence of the magnetization of \MgGd\ shown in Fig.~\ref{fig:static_magnetism}(a) was analyzed
by finding the root of the function $f(M)$\,=\,$M - gJB_{J}(||\vec{H}_{\text{tot}}(M)||,T)$ iteratively with Brent's method. Here, $\vec{H}_{\text{tot}} = \vec{H}_{\text{ext}} + \vec{H}_{\text{DD}} + \vec{H}_{\text{J}}$ is the effective total field  in the Brillouin function comprising the external, dipolar and exchange fields, respectively, and $M$ is the average magnetization per ion in units of $\mu_{\text{B}}$. For Gd$^{3+}$ ions, $J=S=7/2$, and as an initial guess for $M$, $gJB_{J}(||\vec{H}_{\text{ext}}(M)||,T)$ was taken. As a first step, only the dipolar interaction was included to estimate its effect on $M(H)$. With the static bulk magnetization, averages of the form $<S_{\text{i}}>$ or $<\sum_{i} S_{\text{i}}>$ etc. are probed, i.e., all spin fluctuations can be considered to be averaged out. Thus, assuming $\vec{M}$ to be collinear to the external field, only the emergent polarization moment $\vec{\mu}_{\text{pol}}$\,=\,$M^{\text{ion}}\vec{b}$ along the external field direction $\vec{b}$ and with $M^{\text{ion}}\vec{b}$ the magnetization per ion, contributes to the dipolar field $\vec{H}_{\text{DD}}$ in this model. For a sample which is fully spin polarized, the powder average of the dipolar field strength is $||\vec{H}_{\text{DD}}|| \approx 0.7$\,T. The result is shown in Fig.~\ref{fig:Mag_Effective_Field_Only_Dipolar}. Clearly, the dipolar interaction, while it does affect the magnetization curve visibly, is not sufficient to explain the strong deviation of the experimental $M(H)$ curve from the Brillouin function. 
\begin{figure}
		\begin{centering}
			\includegraphics[scale=0.2]{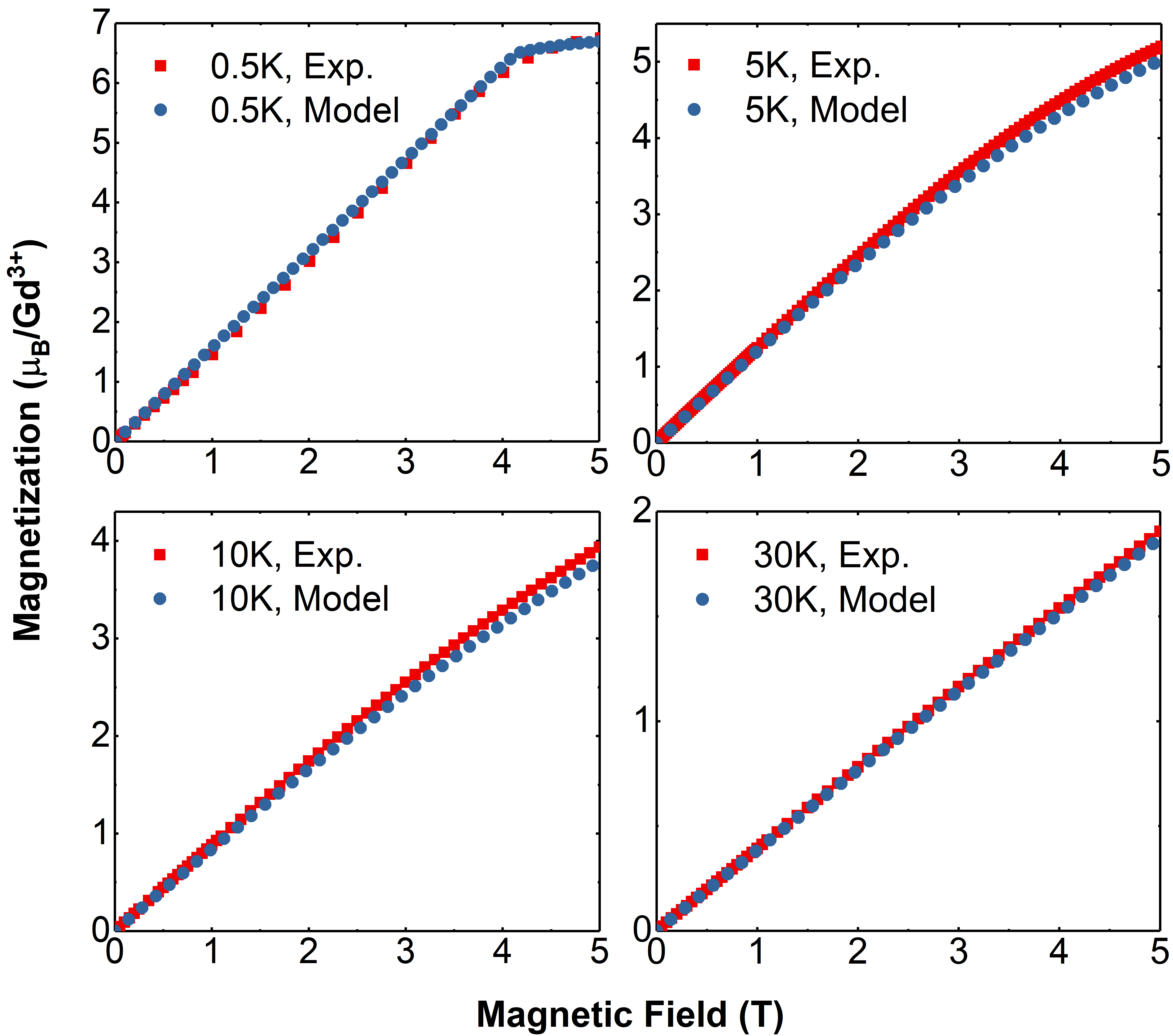}
	\caption{$M(H)$ at different temperatures with the exchange parameter $a_{\text{ex}}$ kept fixed. Red squares represent experimental data, blue dots are the model data.}
		\label{fig:Temp_Dep_M_over_H}
		\end{centering}
	
\end{figure}
%
%

As a next step, the exchange interaction was included in a similar way as the dipolar interaction. The mean exchange field was taken to be $\vec{H}_{\text{J}}$\,=\,$a_{\text{ex}}n_{\text{NN}}M_{i}\vec{b}$ with $a_{\text{ex}}$ an exchange parameter, n$_{\text{NN}}$ the number of nearest neighbors for each ion site, $M_{i}$ the magnetization of the $i$th iteration step and $\vec{b}$ the unit vector in the direction of the external field. This is a valid approach in the case of an isotropic exchange interaction, which is expected for the ensemble of Gd$^{3+}$ isotropic spins with  no contribution from the orbital momentum ($L$\,=\,$0$). In order to translate this value into an exchange constant $J$, the exchange part of the Hamiltonian $\mathcal{H}_{\text{J}}^{\text{j}}$ for spin site $j$ has to be introduced:
\begin{equation}
\mathcal{H}_{\text{J}}^{\text{j}} = J\sum_{i=1}^{n_{\text{NN}}}\vec{S}_{\text{i}}\vec{S}_{\text{j}} = \frac{J}{g^{2}\mu_{\text{B}}^{2}}\sum_{i=1}^{n_{\text{NN}}}\vec{\mu}_{\text{i}}\vec{\mu}_{\text{j}} = -\vec{H}_{\text{J}}\vec{\mu}_{\text{j}}
\end{equation}

\noindent
Note the minus sign here: While, for example, $J$ must be negative for an FM interaction between $\vec{S}_{\text{j}}$ and its surrounding spins, such that the energy gets minimized for parallel $\vec{S}_{\text{j}}$ and $\vec{S}_{\text{i}}$, the exchange field must be \textit{parallel} to $\vec{\mu}_{\text{j}}$ in the case of parallel moments in order to guarantee a (classical) potential energy minimum. The average values are then

\begin{equation}
<\vec{H}_{\text{J}}> = \frac{J}{g^{2}\mu_{\text{B}}^{2}}\sum_{i=1}^{n_{\text{NN}}}<\vec{\mu}_{\text{i}}> =  \frac{n_{\text{NN}}J}{g^{2}\mu_{\text{B}}}\frac{\vec{M}^{\text{ion}}}{\mu_{\text{B}}}
\end{equation}

\noindent
Using $<\vec{H}_{\text{J}}>$\,=\,$a_{\text{ex}}\vec{b}$ and carrying out the fit with $\vec{M}^{\text{ion}}/\mu_{\text{B}}$, it follows $J$\,=\,$g^{2}\mu_{\text{B}}a_{\text{ex}}$.


\subsection{Details of the analysis of the ESR spectral shape}\label{appendix2}

\begin{figure*}
	\centering
	\includegraphics[scale=0.5]{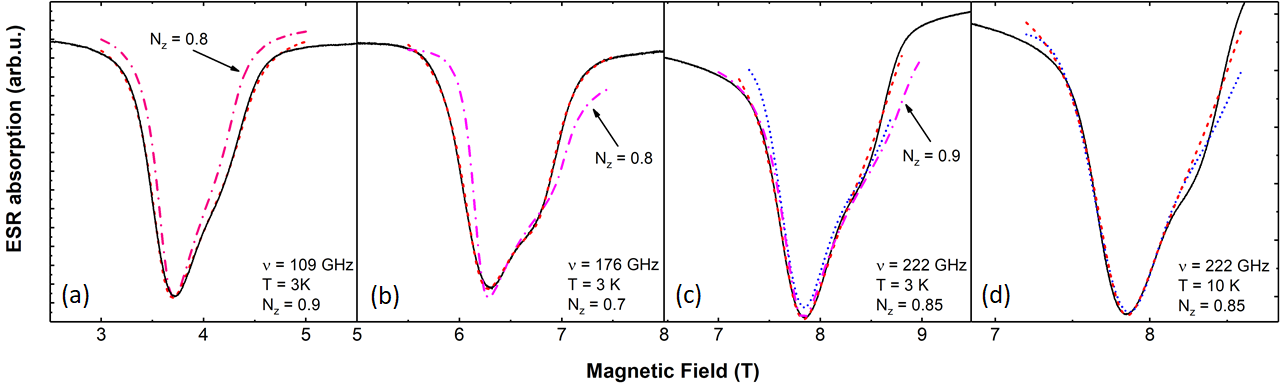}
	\caption{Results of the demagnetization model \cite{Farle1998,Baselgia1988} for the ESR powder spectra. The black solid line shows the experimental data, the red dashed and pink dash-dotted lines the model with a Lorentzian line as basic lineshape, and the blue dotted line the model with a Gaussian line as basic lineshape. The red dashed line represents the Lorentzian with optimal parameters.}
	\label{fig:DemagnetizationResults}
\end{figure*}

\underline{\it Effect of the demagnetizing fields.} The results for the analysis of the influence of the demagnetizing fields on the lineshape of the ESR spectra is shown in Fig.~\ref{fig:DemagnetizationResults}. For details, see the main text. 

\noindent
\begin{figure}[h]
		\includegraphics[scale=0.2]{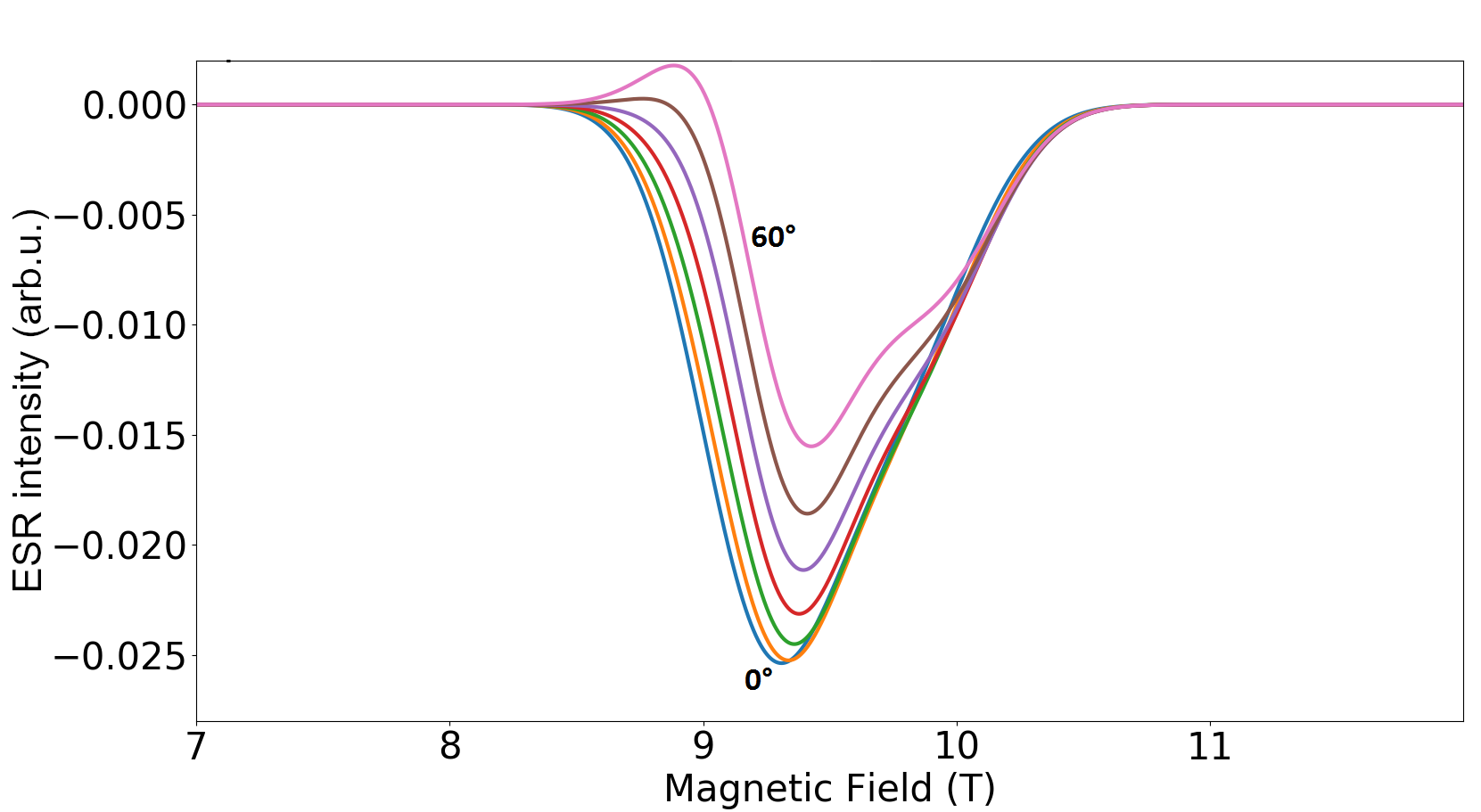}
	\caption{Simulated Gaussian spectra for different mixing angles $\phi = 0-60^\circ$ in Eq.~(\ref{eq:mixing}).}
	\label{fig:mixing}
\end{figure}
\noindent

\underline{\it Mixing of absorption and dispersion.} 
 Due to technical reasons, in the employed high-field ESR setup a mixing of the absorption and dispersion parts (which are related via the Kramers-Kronig relations~\cite{Kendall2003}) cannot always be avoided despite the use of phase-lock detection. Thus the complex signal for the Lorentzian and Gaussian lineshapes reads:
\begin{align}\label{eq:mixing}
L&= \text{cos}(\phi)\frac{A}{1+x^{2}} + \text{sin}(\phi)\frac{Ax}{1+x^{2}}   \\\nonumber
G &= \text{cos}(\phi)A\text{e}^{-x^2/\pi} + \text{sin}(\phi)A\text{e}^{-x^2/\pi}erf(x/\sqrt{\pi})
\end{align} 
\noindent
Here, $L$ is the Lorentzian, $G$ the Gaussian lineshape with mixing angle $\phi$, $erf$ the error function with $x = (H-H_{0})/(\Delta H/2)$, $\Delta H$ = full width at half maximum. The resonance condition requires $h\nu = \text{g}\mu_{\text{B}}H_{0}$ for the peak field $H_{0}$. 
\begin{figure}[h]
	\begin{centering}
		\includegraphics[scale=0.3]{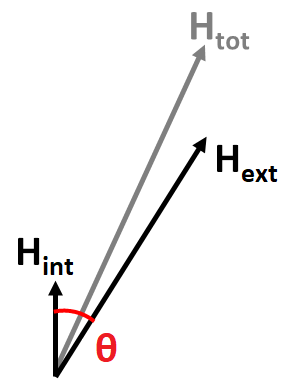}
		\caption{Sketch of the field geometries. Internal field $\boldsymbol{H}_{\text{int}}$ makes an angle $\theta$ with the external field $\boldsymbol{H}_{\text{ext}}$ resulting in an effective total field $\boldsymbol{H}_{\text{tot}}$ according to Eq.~(\ref{totfield}).}
		\label{fig:Internal_External_Field}
	\end{centering}
\end{figure}
\noindent
\noindent
In the fitting procedure such a "mixing" effect was included and its degree was controlled by the respective parameter $\phi$ in Eq.~(\ref{eq:mixing}). Examples of the mixing effect on the ESR spectral shape are shown in Fig.~\ref{fig:mixing}. Even at large $\phi$ the characteristic feature due to the anisotropic internal field can be reliably identified albeit its shape could be affected. The peak field $H_{\rm 0}$ in the fitting procedure is the vector sum of the externally applied field and the internal field, $H_{\rm 0}$\,=\,$||\vec{H}_{\rm ext} + \vec{H}_{\rm int}|| = (H_{\text{int}}^2 + H_{\text{ext}}^2 - 2H_{\text{int}}H_{\text{ext}}\,\text{cos}\theta)^{1/2}$ with $\vec{H}_{\rm int}$ a phenomenological parameter describing the magnitude of the shoulder feature and a powder average taken over all orientations of $\vec{H}_{\rm int}$. The sketch of the field geometries is shown in Fig.~\ref{fig:Internal_External_Field}.

%
%
\begin{figure*}[t]
\includegraphics[scale=0.5]{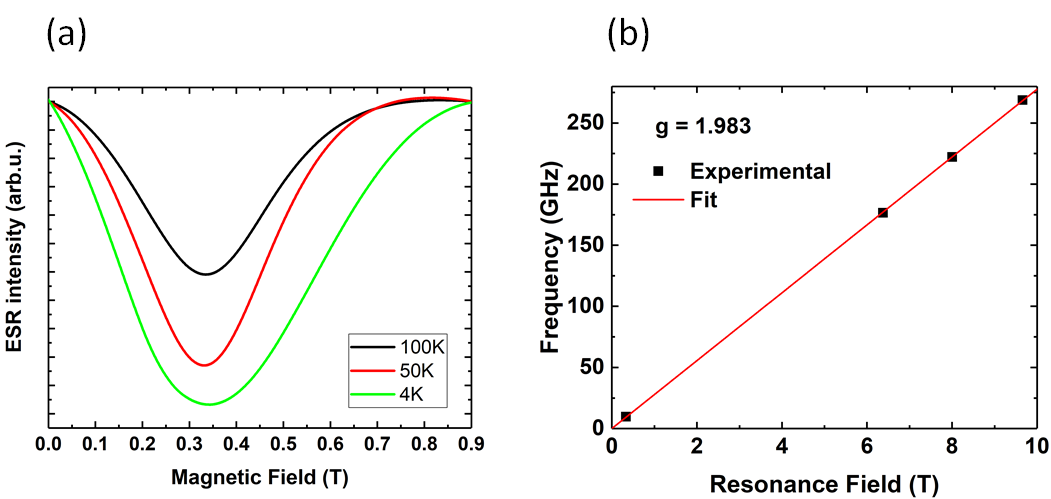}
\caption{(a) Integrals of the recorded absorption derivative curves $dP(u)/dH$ at 10\,GHz (X-band) at different temperatures. (b) g-factor fit at $T$\,=\,40\,K.}
\label{fig:Xband_gfactor40K}
\end{figure*}

\subsection{X-Band data}\label{appendix5}
\begin{figure*}
	\includegraphics[scale=0.28]{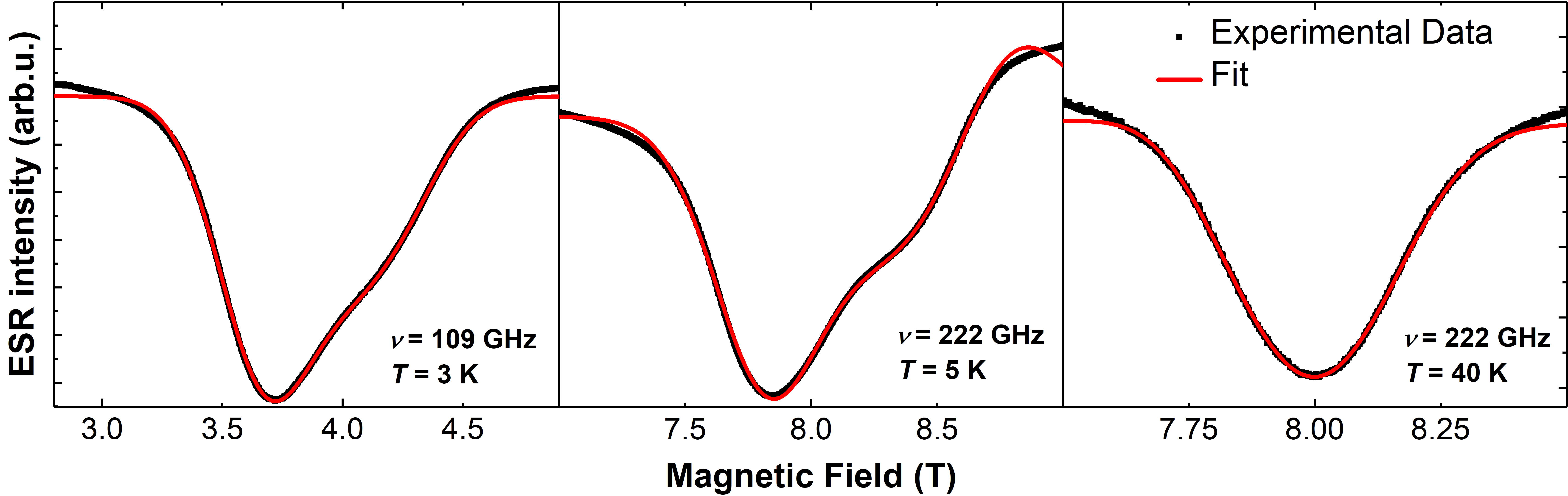}
\caption{Exemplary fitting plots of selected ESR spectra for the internal field model (red lines) in comparison with the experimental data (black symbols) for $\nu = 109$\,GHz at $T=3$\,K and $\nu = 222$\,GHz at $T=5$ and 40\,K.  For details see text.}
\label{fig:Exemplary_Fitting_Plots}
\end{figure*}
In addition to high-field ESR, X-band spectra at a frequency of $\nu = 10$\,GHz were recorded [see Fig.~\ref{fig:Xband_gfactor40K}(a)]. A regular single-peak lineshape without substructure is observable in the temperature range 4\,-\,100\,K. This shows that down to an energy of $h\nu\,=\,0.04$\,meV (with a resonance at a field of $\approx\,0.36$\,T), there is no excitation gap for the resonance. A fit of the X-band and HF-ESR data at $T= 40$\,K to the resonance condition $h\nu = g\mu_{\rm B}H_{\rm res}$ in Fig.~\ref{fig:Xband_gfactor40K}(b) yields the $g$-factor $g = 1.987$.



\end{document}